\documentclass[10pt,a4paper]{article}
\usepackage{inputenc}
\usepackage{setspace}
\usepackage{amsmath}
\usepackage{amsfonts}
\usepackage{amssymb}
\usepackage{amsthm}
\usepackage{bm}
\usepackage{multirow}
\usepackage{graphicx}
\usepackage{enumitem}
\graphicspath{ {./figures/} }
\usepackage{dsfont}
\usepackage[margin=1in]{geometry}
\usepackage[title]{appendix}
\usepackage{authblk}
\usepackage{xcolor}
\usepackage{times}
\usepackage{bm}
\usepackage{natbib}
\usepackage{graphicx}
\usepackage{float}
\usepackage{xcolor}
\usepackage{amsfonts}
\usepackage{bbm}
\usepackage{booktabs} 
\usepackage{caption}
\usepackage{subcaption}
\usepackage{enumitem}
\usepackage{mathrsfs} 
\usepackage{accents}
\usepackage{xr}
\usepackage{hyperref}
\usepackage{optidef}
\usepackage[mathscr]{euscript}
\usepackage{bbm}

\newtheorem{theorem}{Theorem}

\newtheorem{remark}{Remark}

\newcommand{\ind}{\perp\!\!\!\!\perp}

\newcommand{\cZ}{\mathcal{Z}}

\newcommand{\ccC}{\mathcal{C}}
\newcommand{\ccZ}{\mathcal{Z}}
\newcommand{\ccA}{\mathcal{A}}

\newtheorem{assumption}{{Assumption}}
\newtheorem{condition}{{Condition}}
\newtheorem{lemma}{{Lemma}}

\usepackage{caption}
\usepackage[noend]{algpseudocode}

\usepackage{tikz}
\usetikzlibrary{positioning}

\tikzset{
    every neuron/.style={
        circle,
        draw,
        minimum size=1cm
    },
    neuron missing/.style={
        draw=none, 
        scale=4,
        text height=0.333cm,
        execute at begin node=\color{black}$\vdots$
    },
}

\usepackage{natbib}
\title{Semiparametric Mediation Analysis with Separately Observed Mediator and Outcome under Unmeasured Confounding}
\author{Sijia Li\textsuperscript{1*}}
\author{Ruoyu Wang\textsuperscript{2*}}

\affil{\textsuperscript{1}Department of Biostatistics, University of California, Los Angeles}
\affil{\textsuperscript{2}Department of Biostatistics, Harvard T.H. Chan School of Public Health}

\footnotetext[1]{These authors contributed to this work equally.}
\date{}

\begin{document}
\maketitle

\begin{abstract}
Mediation analysis is widely used to disentangle causal pathways, yet in many real-world studies the mediator $M$ and outcome $Y$ are never jointly observed. This incompleteness breaks the standard identification strategy for natural direct and indirect effects. We introduce a novel data fusion framework that restores the identification by combining two incomplete data sources, one measuring $M$ and the other measuring $Y$. Our approach leverages shared instrumental variables (IVs) to circumvent the need to observe $(M,Y)$ jointly, remains valid under unmeasured confounding via a no-interaction condition, and accommodates covariate and exposure shifts across data sources under a latent alignment condition. We establish two identification strategies, one for settings with a known set of valid IVs, and another for settings where valid IVs must be learned. We further develop semiparametric, influence-function-based estimators with multiple robustness properties, and propose an estimator that attains the semiparametric efficiency bound under appropriate conditions. We apply our framework to quantify the extent to which the effect of SNP rs610932 on dementia risk is mediated through immune-related gene-expression pathways.
\end{abstract}

\newpage 
\section{Introduction}
\label{s:intro}
Mediation analysis is a powerful statistical tool for understanding the causal relationship between an exposure $D$, a mediator $M$, and an outcome $Y$. It decomposes the total effect of $D$ on $Y$ into a pathway operating through the intermediate variable $M$ and those not, which are formalized as the natural indirect and direct effect \citep{robins1992identifiability,pearl2022direct}.  This decomposition turns ``does it work" into ``through what pathways does it work", thus facilitating the identification of surrogate outcomes \citep{huang2023surrogate,le2024time} and the underlying diseases mechanisms \citep{vanderweele2012genetic,inzucchi2018does,liu2022large,wanner2024sglt2}, and has exhibited potential to drive intervention designs \citep{mackinnon1994analysis,keele2015identifying} and policy decisions \citep{imai2010general,rudolph2019causal}. 

A key practical challenge in mediation analysis is that the mediator and outcome are frequently measured in different studies, so $(M,Y)$ are never jointly observed. In integrative genomics, gene expression is measured in eQTL studies, while disease status comes from separate GWAS; consequently, widely used pipelines explicitly integrate eQTL and GWAS information to investigate whether genetic associations with disease act through gene regulation \citep{zhu2016integration,gusev2016integrative,porcu2019mendelian}. In population health, 
large observational cohorts are designed to measure exposures at scale and follow participants for health outcomes. In contrast, granular intermediate phenotypes relevant to mediation -- such as longitudinal medication use, biomarkers, and other care processes -- are typically captured as part of routine clinical workflows and therefore reside only in health-care databases or dedicated prospective sub-studies due to cost and logistical burden \citep{pine2016differential,dreyfuss2021high,derkach2024mediation}. 

This is the central motivating setting for our work. Although in many applications no single data source contains the mediator and the outcome jointly, it is often realistic to have two complementary sources -- one with $(D,Y)$ and another with $(D,M)$ available. We view this as a data-fusion problem where the goal is to combine these incomplete heterogeneous data sources to estimate the natural direct and indirect effects. Recent years have seen substantial progress on fusing multiple sources to estimate causal effects \citep{dahabreh2020extending,rudolph2017robust,hu2022semiparametric,rudolph2025improving,yang2025causal} and other target estimands of interest in general \citep{li2023efficient,graham2024towards,lee2025doubly}. A common theme in this literature is to express the target estimand as a functional of several components of the target distribution, with each component observed in at least one data source. The key assumptions then enforce alignment of these components across sources—typically via invariance of certain conditional distributions—so sources can be fused without introducing bias and in general renders efficiency gains. In contrast, identification of causal mediation effects fundamentally depends on the never-observed mediator-outcome relationship, e.g., $Y\mid M,D$. Consequently, the alignment required in this work is inherently latent, which is understudied in the literature and motivates the need to develop new identification and inference schemes.

Compounding these difficulties, unmeasured mediator--outcome confounding remains a major obstacle to identifying causal mediation effects, even when $M$ and $Y$ are jointly observed \citep{ding2016sharp,sun2023semiparametric,miao2023identifying,rudolph2024using}. In our setting, this problem is even more acute because mediator--outcome confounding concerns precisely the joint relationship between $M$ and $Y$, which is never observed. Consequently, standard identification strategies, diagnostic checks, and sensitivity analyses \citep{vanderweele2010bias,ding2016sharp,ohnishi2026sensitivity} become substantially harder to deploy.

\subsection{Contribution}
\label{s:contribution}
In this work, we introduce a novel data fusion identification framework for mediation analysis that simultaneously tackles the two challenges above: the absence of jointly observed $(M,Y)$ and the presence of unmeasured $M$-$Y$ confounding. We consider the setting with multiple exposures $D \in \mathds{R}^d$ that consist of an exposure of interest $T \in \mathds{R}$, a set of instrumental variables (IVs) $Z$, and a set of auxiliary variables $A$ which have a direct effect on $Y$ but are not of primary interest. Inspired by work of \cite{miao2023identifying}, we make use of $Z$ that are observed in both data sources. These IVs have no direct effect on $Y$ except through $M$, while the conditional distribution of the mediator satisfies a completeness condition such that any variability in the mediator $M$ can be captured by variability in $Z$. Subsequently, for functionals involving the counterfactual outcome and mediator under different exposures, we leverage $Z$ to align the mediator distribution under the observed data to its counterfactual distribution, thereby circumventing the need to ever jointly observe $(M,Y)$. We further consider two scenarios where the researcher knows which components of $D$ are valid IVs, and when the researcher only knows that valid IVs lie within a set of candidate variables but does not know which ones. For each scenario, we propose a novel identification scheme under a no-interaction condition and a cross-source latent alignment condition where the conditional distributions of the outcome, mediator and unmeasured confounders given exposures and covariates are the same, while allowing for covariate and exposure shifts across the two data sources. Such alignment is latent in the sense that the aligning conditional distribution involves variables that are never jointly observed.

Using tools from semiparametric efficiency theory, we develop multiply robust estimators of the natural direct and indirect effects that are asymptotically linear and normal under appropriate conditions. Under a mutual completeness condition, we show these estimators attain the semiparametric efficiency bound. We further derive the canonical gradients for these effects under a weaker completeness condition in the absence of unmeasured confounding. In addition, we draw the connection to the two-sample instrumental variable literature and show that these problems arise as special cases of our setting despite of some technical differences.

\subsection{Related Work}
\label{s: related work}
Our work builds on the growing literature on data fusion. A recent work by \cite{derkach2024integrating} specifically considered integrating data sources with each containing only two of the three key variables, and relied on a parametric form of the distribution of $Y\mid M,D$. Another recent line of semiparametric data fusion literature develops principled ways to combine heterogeneous data sources without imposing parametric assumptions. Although no work from this line has been proposed to specifically estimate the natural direct and indirect effects (to our knowledge), \cite{li2023efficient} and \cite{graham2024towards} both provide a general recipe that can be specialized to mediation estimands. Relatedly, \cite{huang2023surrogate} and \cite{kallus2025role} studied causal effects using short-term surrogate markers, which are also intermediate variables though not necessarily mediators. More broadly, the idea of combining complementary samples with different components of a larger variable vector has also been studied in missing data \citep{robins1994estimation,tsiatis2006semiparametric,wang2009causal} and the semi-supervised learning literature under blockwise missingness \citep{xu2025unified,huang2025efficient,xu2025blockwise}.
While these results are conceptually related to our “missing mediator vs. missing outcome” structure, these aforementioned approaches rely on some overlap sample where the mediator and the outcome are both observed. One exception is the work by \cite{evans2021doubly}, which proposed semiparametric estimators for parametric inferences under a regression setting when the outcome and two covariates of interest are never jointly observed. 

IVs have been used in the literature to infer mediation effects under unmeasured confounding \citep{imai2013experimental,frolich2017direct,rudolph2021complier,rudolph2024using}. Existing works require joint observation of IVs, exposures, mediators, and outcomes and identify mediation effects only for specific compliance groups. On the other hand, the population-level causal effect can be identified using IVs under unmeasured confounding when estimating the average treatment effect of a binary exposure \citep{wang2018bounded,dong2025marginal,chen2025identification}. However, these methods also rely on the joint observation of all variables. Our general framework encompasses two-sample extensions of these approaches as special cases. See Remark~\ref{remark: connect two-sample IV} for further discussion of these connections.

\subsection{Notation and Setup}
\label{s: notations}
We let $Y$, $M$, $X$, and $U$ denote the outcome, mediator, observed confounders, and unmeasured confounders. Let $D:= (T,Z,A) \in \mathds{R}^d$ be a vector of exogenous variables. Among these, $T \in \{0, 1\}$ is the target exposure of interest, $Z \in \mathds{R}^{|\mathcal{Z}|}$ is the subvector of IVs that do not have a direct effect on $Y$, and $A \in \mathds{R}^{|\mathcal{A}|}$ is the subvector of auxiliary variables which have direct effects on $Y$ but their effects are not of primary interest. We use $\ccZ$ and $\ccA$ to denote the index sets corresponding to the components of $Z$ and $A$, and write $|\mathcal{C}|$ for the cardinality of a set $\mathcal{C}$. Let $S$ denote the data source indicator. In an ideal setting, we observe the full data structure $O^F = (S,X,U,D,M,Y) \sim P^F \in \mathcal{P}^F$. Instead, we only observe $(Y,D,X)$ in data source $S=0$ and  observe $(M,D,X)$ in data source $S=1$. The observed data unit can therefore be written as $O = (S,X,D,SM,(1-S)Y)\sim P$. Throughout we use capital letters to denote random variables and the corresponding lowercase letters to denote their realizations.

Conditional laws involving latent or jointly unobserved quantities, such as $U$ or $(M,Y)$, are taken under $P^F$; laws involving only observed quantities are taken under $P$. 
The subscript of $E$ indicates the distribution with respect to which the expectation is taken; when the expectation is taken under $P^F$, we omit the subscript. We slightly abuse the notation by writing distributions and expectations conditional on a lowercase letter to indicate conditioning on a specific value. For example,  we let $P_{V_1 \mid v_2}$ denote the conditional distribution of $V_1 \mid V_2 = v_2$,  $p_{V_1\mid V_2}(v_1\mid v_2)$ denote the corresponding density evaluated at $V_1 = v_1 \mid V_2 = v_2$, and $E[V_1 \mid v_2 ]=E[V_1 \mid V_2 = v_2 ]$.  Here and throughout, we suppose sufficient regularity conditions, such as conditional distributions are well defined, share a common measurable space, and the densities are bounded and square integrable. 

 We aim to estimate the causal mediation effects of the target exposure $T$. For ease of presentation, we consider a binary $T \in \{0,1\}$, although the proposed methods extend readily to discrete multi-level exposure $T$. Following the potential outcome framework \citep{neyman1923application,rubin1974estimating}, we use the nested potential outcome $Y(t,M(t'))$ to denote the hypothetical outcome if the exposure were set at $T=t$ and the mediator were set at what would happen under $T=t'$. For each population $s \in \{0, 1\}$, our goal is to estimate the natural direct effect
\[
   {\rm NDE}_s = E[Y(1, M(0))\mid S = s] - E[Y(0, M(0))\mid S = s],
\]
which measures the effect of the exposure $T$ on the outcome $Y$ if the mediator $M$ were set to the natural value $M(0)$ under no exposure. And the natural indirect effect is
\[
   {\rm NIE}_s = E[Y(1, M(1))\mid S = s] - E[Y(1, M(0))\mid S = s],
\]
which measures the effect of the exposure $T$ through changing the mediator $M$ if the exposure itself $T$ were set at $T = 1$. 
Under our setting, the challenge is to identify the cross-term $E[Y(1, M(0))\mid S = s]$, as $Y$ and $M$ are never jointly observed in the same dataset.


\section{Identification Using IVs}
\label{s: identification}
\begin{figure}[h]
\centering
\includegraphics[scale=1.2]{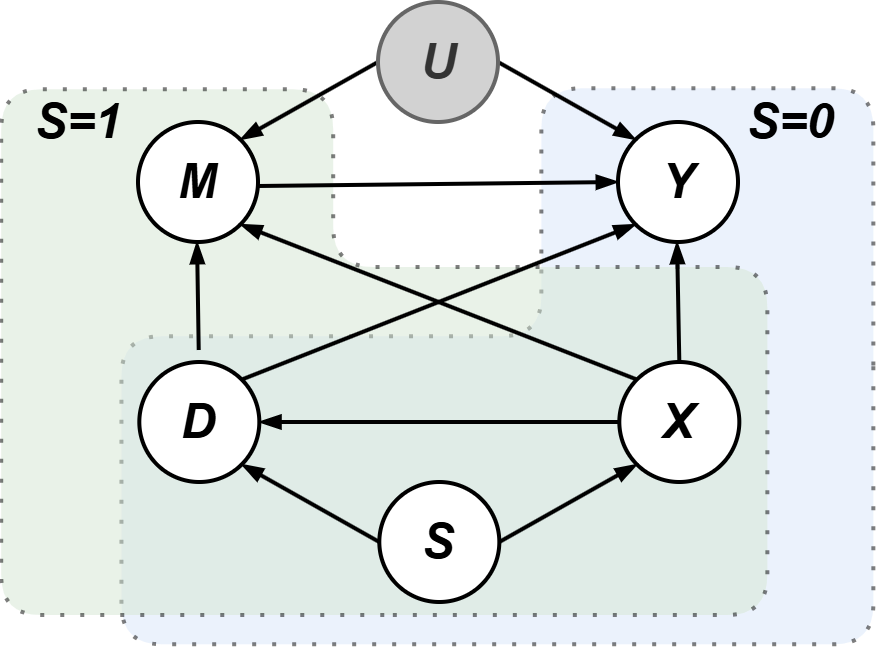}
\caption{An illustration of the data generating process. Dotted polygons indicate the variables jointly observed from each source.
}
\label{fig:dgm}
\end{figure}

We formalize the data generating process through a structural causal model (SCM) \citep{pearl2000models}. We assume that there exists deterministic functions $f_M$ and $f_Y$ such that 
\begin{equation}\label{eq: structural}
    \begin{aligned}
    M & =  f_{M}(T, Z, A, X,  U, \varepsilon_{M}),\\
    Y & = f_{Y}(M, T, A, X, U, \varepsilon_{Y}),
\end{aligned}
\end{equation}
where $\varepsilon_{Y}$ and $\varepsilon_{M}$ are exogenous error terms. For any exposure levels $t$ and $t^{\prime}$, the potential mediators and outcomes can be defined as $M(t) = f_{M}(t, Z, A, X, U, \varepsilon_{M})$ and $Y(t, M(t^{\prime})) = f_{Y}(M(t^{\prime}), t, A, X, U, \varepsilon_{Y})$, where $T$ is removed from the structural model and assigned as a fixed value. Figure~\ref{fig:dgm} illustrates the data generating process.

Our identification strategies rest on the following assumptions. Assumption~\ref{ass: unconfound exposure} rules out exposure-confounder, exposure-outcome and exposure-mediator confounding once adjusted for the observed covariates $X$ for each population $s\in \{0,1\}$. Assumption~\ref{ass: unconfound mediator} rules out other latent mediator-outcome confounding once adjusted for $(U,X,S)$. Under the structural causal model \eqref{eq: structural} with $U = \emptyset$, Assumptions~\ref{ass: unconfound exposure} to \ref{ass: positivity} recover the ignorability component of the standard sequential ignorability conditions in \cite{imai2010general}, together with positivity of the exposure. We do not impose a separate mediator positivity as the corresponding identifying role is replaced by the bridge-function conditions introduced later.
\begin{assumption}[Unconfounded exposure]\label{ass: unconfound exposure}
    $D \ind (\varepsilon_{Y}, \varepsilon_{M}, U)\mid X, S$.
\end{assumption}
\begin{assumption}[Latent unconfounded mediator]\label{ass: unconfound mediator}
    $\varepsilon_{Y} \ind \varepsilon_{M}\mid X, U, S$.
\end{assumption}
\begin{assumption}[Positivity of $T$]\label{ass: positivity}
    There exists a constant $\epsilon>0$ such that $\epsilon< P(T=1 \mid z,a,x,s) < 1 - \epsilon$ for any $z,a,x,s$ $P$-everywhere.  
\end{assumption}
The following assumption is imposed to accommodate the presence of an unmeasured confounder $U$.
\begin{assumption}[No $M$-$U$ interaction]\label{ass: no interaction}
    $E[Y\mid M = m, D, X, U,S] - E[Y\mid M = m^{\prime}, D, X, U,S]$ does not depend on $U$ for any $m$ and $m^{\prime}$. 
\end{assumption}

Assumption \ref{ass: no interaction} states that,  conditional on $(D,X,S)$, $U$ is not an effect modifier. This assumption is weaker than the cross-world independence condition in the standard mediation literature, 
which is typically justified by assuming  $U=\emptyset$. While Assumptions~\ref{ass: unconfound exposure} to \ref{ass: no interaction} are sufficient for the identification of mediation effects under the structural equation \eqref{eq: structural} and a single data source with $(X,D,M,Y)$ jointly observed, they do not resolve our issue as $M$ and $Y$ are never observed together. To address this, we impose the following assumption and condition:
\begin{assumption}[Cross-source alignment]
There exists $c>0$ such that $c<P(S=1\mid D,X)<1-c \quad P\text{-a.s.}$ In addition, $S\ind (U,M,Y)\mid D,X$. 

\label{ass: transportability}
\end{assumption}
Assumption~\ref{ass: transportability} resembles the overlap and missing-at-random conditions used in existing data fusion literature: after adjusting for appropriate effect modifiers, the conditional distributions of $U$, $M$ and $Y$ are exchangeable across data sources. As these conditions are only required on the conditional distributions of selected variables, such framework can thus accommodate heterogeneous joint distributions across the two sources under covariate $X$ and exposure $D$ shifts. Lastly, we invoke the following completeness condition to establish identification.
\begin{condition}[$M-Z$ Completeness]\label{cond: med completeness} 
    For any $s$, $t$, $a$, $x$, and any square-integrable function $h$, $E[h(M)\mid T = t, Z, A = a, X = x,S=s] = 0$ almost surely implies $h = 0$ almost surely. 
\end{condition}
Condition~\ref{cond: med completeness} states that, conditional on $(T,A,X,S)$, any variability in $M$ is captured by variability in $Z$. Completeness holds for very general exponential families of distributions \citep{lehmann2011completeness,newey2003instrumental,d2011completeness}, and is commonly satisfied for continuous $Z$ and $M$ under a mild regularity condition \citep{chen2014local,andrews2011examples}. \cite{miao2023identifying} and \cite{guo2025comparing} give a great discussion on the role of completeness in recent nonparametric and semiparametric identification problems and we direct readers' attention  there for a detailed overview. Different from the conventional data combination setting discussed in Section 4.2.2 of \cite{ridder2007econometrics}, the completeness condition here does not directly identify the joint distribution of $Y, M, D, X, S$ because $Y\not\ind Z \mid M, T, A, X, S$ in the presence of the unmeasured confounder $U$. Fortunately, under Condition~\ref{cond: med completeness}, we can leverage $Z$ to reweigh the observed $M$ strata to match the corresponding counterfactual strata involved in the cross-terms, as formulated in the following lemma. The weights enable us to identify NDE and NIE without identifying the joint distribution.
\begin{lemma}[Existence of bridge functions $w_s$] \label{lem: w} Under Condition~\ref{cond: med completeness} and the regularity Condition S1 in Supplementary Material, for $s \in \{0, 1\}$, there exists some function vector $w_s = (w^{11}_s,w^{10}_s,w^{00}_s)$ that satisfies
\begin{align}
    &\int w^{tt'}_{s}(z, a, x)p_{M\mid T, Z,A,X, S}(m\mid t, z, a, x, 1) p_{Z, T\mid A, X, S}(z, t\mid a, x, 0)dz \nonumber\\
    & = \int p_{M\mid T, Z,A,X, S}(m\mid t', z, a, x, 1)p_{Z\mid A, X, S}(z\mid a, x,  s)dz \label{eq: lemma1}
\end{align}
for $t, t' \in \{0,1\} $ and $a, x$ $P$-everywhere.
\end{lemma}
The bridge function $w_{s}^{tt'}$ reweighs the mediator strata $M, T=t\mid A,X,S=0$ such that it aligns with $M(t') \mid A,X,S=s$.  When $t=t'$, equation~\eqref{eq: lemma1} admits the simple solutions with $w^{tt}_s(z,a,x) = p_{Z\mid A,X,S}(z\mid a,x,s)/\{P(T=t\mid z,a,x,S=0)p_{Z\mid A,X,S}(z\mid a,x,0)\}$.  More generally, however, solutions to \eqref{eq: lemma1} are not unique as uniqueness holds only if $Z$ is also complete in $M$. Importantly, this lack of uniqueness is harmless for identification as all valid bridge functions $w_s$ will lead to the same reweighted counterfactual mediator distribution. The integral equation \eqref{eq: lemma1} can be solved by solving a finite-dimensional system of linear equations when $M$ and $Z$ are discrete. Solutions of explicit forms can be available even when $M$ and $Z$ are continuous, provided that they follow certain parametric distributions. See Supplementary Material Section S8.3 for examples.

\begin{theorem}[Identification with known IVs] \label{thm:identification}
Suppose $E_P[\{w_s^{tt'}(Z,A,X)\}^2] < \infty$ for $s \in \{0,1\}$ and $t,t'\in \{0,1\}$. Let $\mathcal{P}$ denote the set of observed data distributions induced by full-data laws $P^F \in \mathcal{P}^F$ satisfying Assumptions \ref{ass: unconfound exposure}--\ref{ass: transportability} and Condition \ref{cond: med completeness}. For any $P \in \mathcal{P}$, ${\rm NDE}_s$ and ${\rm NIE}_s$ can be identified as $\psi^{\rm NDE}_s(P)  = \psi^{10}_s(P)-\psi^{00}_s(P)$ and $\psi^{\rm NIE}_s(P)  = \psi^{11}_s(P) -  \psi^{10}_s(P)$, where $\psi^{11}_s$, $\psi^{10}_s$, $\psi^{00}_s$: $\mathcal{P} \rightarrow \mathbb{R}$ and
\[\begin{aligned}
    \psi^{11}_s(P) & :=  E_P\left[E_P[Y\mid T = 1, Z, A, X, S = 0]\mid S = s\right]\\
    \psi^{10}_s(P) & :=   E_P\left[E_P\left[\mathds{1}(T = 1)w^{10}_{s}(Z, A, X)Y\mid  A, X,  S = 0\right]\mid S = s\right] \\
    \psi^{00}_s(P) & :=  E_P\left[E_P[Y\mid T = 0, Z, A, X, S = 0]\mid S = s\right].
\end{aligned}
\]
\end{theorem}

When the index set of IVs $\ccZ$ is unknown, it is still possible to identify a subset sufficient for identification under additional assumptions. We reduce the search space by restricting attention to variables that are informative about the mediator.  For $j \in \{1, \dots, d\}$, let $D_{j}$ denote the $j$-th component of $D$ and $D_{-j}$ the subvector of $D$ that excludes $D_{j}$. We define $\ccC := \{j: D_{j} \not\ind_P M \mid D_{-j}, X, S \text{ and } D_j\ \text{is not the target exposure}\  T\} $ the set of candidate IVs that provide additional information about $M$ beyond exposures $D_{-j}$ and covariates $X$. In practice, $\ccC$ can be viewed as a reduced candidate set obtained from a preliminary screening step, and may consist of some or all components of $\ccA$, and some or all components $\ccZ$.  This reduction is without loss for identification as variables outside $\ccC$ do not contribute to the completeness requirement in Condition \ref{cond: med completeness}. Specifically, Condition \ref{cond: med completeness} holds for $Z$ if and only if it holds for the subvector of $Z$ indexed by $\ccZ \cap \ccC$. Therefore, once $\ccC$ is formed, it suffices to identify the effective valid IV set $\ccZ \cap \ccC = \ccC \setminus \ccA$ or the set $\ccA \cap \ccC$. We establish the identification result under the following assumption.

\begin{assumption}[Sparsity]\label{ass: sparsity}
    $|\ccA\cap\ccC| \leq (|\ccC| - 1) / 2$.
\end{assumption}

\begin{condition}[$M-D_j$ Completeness]\label{cond: strong completeness}
    For $j \in \ccC$, any $s$, $d_{-j}$, $x$ and any square-integrable function $h$, $E[h(M)\mid D_{j}, D_{-j} = d_{-j}, X = x, S=s] = 0$ almost surely implies $h = 0$ almost surely.
\end{condition}
\begin{condition}\label{cond: non-trivial dependence}
    For some $u_{*}$ in the support of $U$,
    $ E[Y\mid T, M, A, X, U = u_{*}, S ] +E[Y\mid  T, A, X, S ] - E[E[Y\mid T, M, A, X, U = u_{*}, S ]\mid T,A,X,S]$ depends on each component in $\ccA \cap \ccC$.
\end{condition}
Assumption \ref{ass: sparsity} requires more than half of the candidate IVs to be valid. Condition \ref{cond: strong completeness} strengthens Condition \ref{cond: med completeness} by requiring the same type of completeness to hold for each $D_j$, $j\in\ccC$. If $\ccZ\cap\ccC\neq\emptyset$, it implies Condition \ref{cond: med completeness}. Condition \ref{cond: non-trivial dependence} 
requires that each invalid candidate IV in $\ccA \cap \ccC$ remains detectable in the bridge function induced by the observed-data quantity $E_P[Y\mid D,X,S=0]$; that is, it shows up in the function $r_*(M,T,A,X)$ such $E_P[r_*(M,T,A,X) \mid D,X,S=1] = E_P[Y\mid D,X,S=0]$. We are now ready to establish the identification of the subset $\cZ \cap \ccC$.

\begin{theorem}[Identification of $\ccZ \cap \ccC$]\label{thm:identification IV set}
Under Assumptions \ref{ass: unconfound exposure} -- \ref{ass: sparsity} and Conditions \ref{cond: strong completeness} and \ref{cond: non-trivial dependence}, when $\ccZ\cap \ccC \neq \emptyset$, there exists a unique function $r_{*}(M,D,X)$ that depends on no more than $(|\ccC| - 1) / 2$ variables in $\{D_{j}\}_{j\in \ccC}$ that satisfies $E_P[Y\mid D, X, S = 0] = E_P[r_{*}(M, D, X)\mid D, X, S = 1]$. The index set $\ccZ\cap\ccC$ is identified as $$\ccZ\cap\ccC= \{j\in \ccC: r_{*}(M, D, X)\ \text{does not depend on}\ D_{j}\}.$$
\end{theorem}
Theorem \ref{thm:identification IV set} implies that $r_{*}$ is the most sparse solution to $E_P[Y\mid D, X, S = 0] = E_P[r_{*}(M, D, X)\mid D, X, S = 1]$ in the sense that it depends on the smallest number of components in $\ccC$ among all possible solutions.
The identification of ${\rm NDE}_s$ and ${\rm NIE}_s$ then follows exactly as in Theorem~\ref{thm:identification}, with $Z$ replaced by the identified valid IV subvector indexed by $\ccZ \cap \ccC$. For practical implementation, we propose a reproducing kernel-based nonparametric IV selection algorithm in Supplementary Section S7 which selects IVs leveraging the identification formulation in Theorem \ref{thm:identification IV set}.

\section{Methods}
\label{s: method}
To construct efficient and robust estimators in the two-sample setting with minimal assumptions on $\mathcal{P}$, we resort to tools from semiparametric efficiency theory \citep{bickel1993efficient}. The construction of such estimators requires a key object, namely gradients, and we defer the reader's attention to Supplementary Section S2 for an overview and to 
\cite{pfanzagl1985contributions,van2000asymptotic} for a more detailed read. Hereafter, we focus on estimating the natural direct and indirect effect for the population $S=0$ for clarity, and defer the results for $S=1$ to the Supplementary Section S5. In the main text, we write $w^{tt'}$ for $w^{tt'}_{0}$ for simplicity. Henceforth we suppress the superscript $s$. For ease of notation, we let $G:= (A,X)$, $\mu_t(Z,G): = E_P[Y\mid T=t, Z, G, S=0]$, $\nu_t(M,G)$ be the solution to ${\mathds{M}}_{\nu_t}(t,Z,G) = \mu_t(Z,G)$, where we define ${\mathds{M}}_{\nu_t}(t',Z,G): = E_{P}[\nu_t(M,G)\mid T=t',Z,G,S=1]$. We let $\mu_{10}(Z,G): = {\mathds{M}}_{\nu_1}(0,Z,G)$, and $\lambda(T,Z,G): = p(T,Z,G \mid S=0)/p(T,Z,G \mid S=1)$. We derive the gradients for mediation effects in the following lemma.

\begin{lemma}[A set of gradients for $\psi^{\rm NDE}$ and $\psi^{\rm NIE}$]\label{lemma: IF}
    Under Assumptions~\ref{ass: unconfound exposure} to \ref{ass: transportability} and Condition~\ref{cond: med completeness}, a set of gradients of $\psi^{11}$, $\psi^{10}$ and $\psi^{00}$ relative to $\mathcal{P}$ at $P$ are:
        \begin{align*}
        D^{11}_P(o) & = \frac{\mathds{1}(s=0)}{P(S=0)}\left\{\mathds{1}(t=1) w^{11}(z,g)\left(y - \mu_1(z,g)\right) + \mu_1(z,g) - \psi^{11}(P)\right\}\\
       D^{10}_{P}(o)   & = \frac{\mathds{1}(s=0)}{P(S=0)}\left\{\mathds{1}(t=1) w^{10}(z,g) \left(y - \mu_1(z,g)\right) + \mu_{10}(z,g) - \psi^{10}(P)\right\} \\
        & \quad + \frac{\mathds{1}(s=1)}{P(S=1)}\mathds{1}(t=0)w^{00}(z,g)  \lambda(t,z,g) \left(\nu_1(m,g) - \mu_{10}(z,g)\right)   \\
        & \quad - \frac{\mathds{1}(s=1)}{P(S=1)}\mathds{1}(t=1)  \lambda(t,z,g) w^{10}(z,g)\left(\nu_1(m,g) - \mu_1(z,g)\right) \\
         D^{00}_P(o) & = \frac{\mathds{1}(s=0)}{P(S=0)}\left\{\mathds{1}(t=0)w^{00}(z,g)\left(y - \mu_0(z,g)\right) + \mu_0(z,g) - \psi^{00}(P)\right\}.
    \end{align*}
    Therefore, the gradients for $\psi^{\rm NDE}$ and $\psi^{\rm NIE}$ relative to $\mathcal{P}$ at $P$ are $D^{\rm NDE}_P(o) = D^{10}_{P}(o) - D^{00}_{P}(o)$ and $ D^{\rm NIE}_P(o) = D^{11}_{P}(o) - D^{10}_{P}(o)$.

\end{lemma}
In Lemma~\ref{lemma: IF}, the gradients for $\psi^{11}(P)$ and $\psi^{00}(P)$ depend only on the $S=0$ sample, as these two components are identifiable from the source $S=0$ alone. In contrast, the cross-term $\psi^{10}(P)$ depends on the conditional outcome distribution under $T=1$ and mediator distribution under $T=0$, which are not jointly observed in any single source. Its gradient therefore leverages the cross-term bridge function $w^{10}$ in addition to $w^{00}$. Notably, Lemma~\ref{lemma: IF} introduces a family of valid gradients indexed by the choices of $w$, as \eqref{eq: lemma1} may admit multiple solutions. 
Examining the analytical form of the variance of the gradients, the variance-optimal choice corresponds to the one with minimum weighted $L^0_2(P)$-norm, with weights determined by the conditional variances of the outcome and mediator components across data sources. 

 While the gradients $D_P^{\rm NDE}$ and $D_P^{\rm NIE}$ do not live in the tangent space of $\mathcal{P}$, they do characterize the semiparametric efficiency bound (and therefore are the canonical gradients for mediation effects) for another model described below.
\begin{condition}[$Z-M$ Completeness]\label{cond: iv completeness}
    For any $t$, $g$, and any square-integrable function $h$, $E[h(Z)\mid T = t, M, G = g,  S=0] = 0$ almost surely implies $h = 0$ almost surely. 
\end{condition}

\begin{condition}[Common mediator support] \label{cond: common support}
    For $P_{T,Z,G \mid S=1}$-almost every $(t,z,g)$, and $u$ in a set of positive probability under $P^F_{U\mid X=x}$, $P^F_{M\mid T=t, Z=z, G=g, U = u, S=1} \sim P^F_{M\mid T=t, Z=z, G=g, S=1}$, where $P \sim Q$ denotes mutual absolute continuity.
\end{condition}

\begin{lemma}[Characterizing the tangent space of $\mathcal{P}_{\rm comp}$] \label{lem:tangent_space}
Let $\mathcal{P}_{\rm comp} \subset \mathcal{P}$ denote the set of observed data distributions induced by full data distributions $P^F$ that satisfy Assumptions~\ref{ass: unconfound exposure}-\ref{ass: transportability} and Conditions~\ref{cond: med completeness}, \ref{cond: iv completeness} and \ref{cond: common support}. Then, $\mathcal{P}_{\rm comp}$ is locally nonparametric at $P$. That is, $ \mathcal{T}(P,\mathcal{P}_{\rm comp}) = L_2^0(P)$ where $L_2^0(P)$ is the set of mean zero square integrable functions under $P$.
\end{lemma}
To build intuition, although $\mathcal{P}_{\rm comp}$ encodes structural restrictions -- most notably the conditional independence $Z \perp Y \mid M,T,G,U,S$ and the latent alignment of the two sources $(Y,M,U) \perp S \mid D,X$ -- these constraints do not shrink the tangent space of  $\mathcal{P}_{\rm comp}$ when Conditions \ref{cond: iv completeness} and \ref{cond: common support} hold. 
This picture breaks down once $Z$ is no longer conditionally complete in $M$. For such a case, we characterize the tangent space when $U = \emptyset$ and derive the canonical gradients of the natural direct and indirect effect in Section~\ref{s: extension}.

\begin{theorem}[Mutual completeness implies efficiency] \label{thm: mutual complete}
When $P \in \mathcal{P}_{\rm comp}$, the bridge functions admit a unique solution with $w^{11}(z,g) = 1/\pi(z,g)$, $w^{00}(z,g) = 1/(1-\pi(z,g))$ with $\pi(z,g) = P(T=1\mid Z=z, G=g, S=0)$ and the remaining $w^{10}$ being the unique solution to its corresponding integral equation. Moreover, $D_{P}^{\mathrm{NDE}},D_{P}^{\mathrm{NIE}}$ are the only, and thus canonical, gradients of $\psi^{\rm NDE}$ and $\psi^{\rm NIE}$ relative to $\mathcal P_{\rm comp}$ at $P$.
\end{theorem}
For the estimation of $\psi^{11}(P)$ and $\psi^{00}(P)$ under Assumption \ref{cond: iv completeness}, our findings are consistent with those of \cite{kallus2025role}, which finds that auxiliary data containing only a surrogate (mediator) does not yield any efficiency gains without further assumptions. In practice, mutual completeness implies that neither $Z$ nor $M$ loses information about the other, and is only possible when $Z$ and $M$ have the same dimension.

\section{Estimation and Asymptotic Properties}
\label{s: estimation}
We propose to use the derived gradients $D_P^{\rm NDE}$ and $D_P^{\rm NIE}$ to construct one-step estimators \citep{bickel1993efficient}. Given an initial estimate $\widehat{P}$ of $P$, a one-step estimator takes the form of $\widehat{\psi} = \psi(\widehat{P}) + \frac{1}{n}\sum_{i=1}^n D_{\widehat{P}}(O_i)$, where $\psi(\widehat{P})$ is the plug-in estimator and $D_{\widehat{P}}$ is a gradient of $\psi$ evaluated at $\widehat P$. This construction is motivated by the idea of  de-biasing the plug-in estimator via its first-order von Mises expansion, and can be viewed as a semiparametric analogue of taking a single Newton–Raphson step toward the maximum likelihood estimator in parametric models \citep{pfanzagl1985contributions}. Under appropriate conditions, this estimator will be consistent and asymptotically normal with asymptotic variance given by the variance of $D_P$, thus facilitating the construction of confidence intervals.

To construct one-step estimators for $\psi^{\rm NDE}(P)$ and $\psi^{\rm NIE}(P)$, we need to estimate components of $P$ that are needed for the evaluation of $D_P^{\rm NDE}$ and $D_P^{\rm NIE}$. Specifically, the list of nuisance parameters includes  $ (\mathds{M}_\nu,\nu, \lambda,  w)$, where we define ${\mathds{M}}_\nu:= ({\mathds{M}}_{\nu_0},{\mathds{M}}_{\nu_1})$ and $\nu := (\nu_0,\nu_1)$. Under a generic $\widehat P$, we denote the corresponding estimates as $(\widehat{\mathds{M}}_\nu,\widehat \nu, \widehat \lambda,  \widehat w)$. Let $\mathds{P}_n Q$ denote the empirical mean of a generic function $Q$. Then the one-step estimators are given by $\widehat{\Psi}^{\rm NDE}  = \widehat{\psi}^{\rm 10} - \widehat{\psi}^{\rm 00}$ and $ \widehat{\Psi}^{\rm NIE}  = \widehat{\psi}^{\rm 11} - \widehat{\psi}^{\rm 10}$  where
\begin{align*}
   \widehat{\psi}^{\rm 11} & = \mathds{P}_n \left[\frac{\mathds{1}(S=0)}{\widehat{P}(S=0)}\left\{\mathds{1}(T=1)\widehat w^{11}(Z,G)\left(Y - \widehat{\mathds{M}}_{\widehat \nu_1}(1,Z,G)\right) + \widehat{\mathds{M}}_{\widehat \nu_1}(1,Z,G) \right\}\right] \\
   \widehat{\psi}^{\rm 10}& =  \mathds{P}_n \bigg[\frac{\mathds{1}(S=0)}{\widehat P(S=0)}\left\{\mathds{1}(T=1) \widehat w^{10}(Z,G) \left(Y - \widehat{\mathds{M}}_{\widehat \nu_1}(1,Z,G)\right) + \widehat{\mathds{M}}_{\widehat \nu_1}(0,Z,G) \right\} \\
    & \qquad + \frac{\mathds{1}(S=1)}{\widehat P(S=1)}\mathds{1}(T=0) \widehat w^{00}(Z,G)  \widehat \lambda(T,Z,G) \left(\widehat \nu_1(M,G) - \widehat{\mathds{M}}_{\widehat \nu_1}(0,Z,G)\right)   \\
    & \qquad - \frac{\mathds{1}(S=1)}{\widehat P(S=1)}\mathds{1}(T=1)  \widehat \lambda(T,Z,G) \widehat w^{10}(Z,G)\left(\widehat \nu_1(M,G) - \widehat{\mathds{M}}_{\widehat \nu_1}(1,Z,G)\right) \bigg] \\
    \widehat{\psi}^{\rm 00} & = \mathds{P}_n \left[\frac{\mathds{1}(S=0)}{\widehat{P}(S=0)}\left\{\mathds{1}(T=0) \widehat w^{00}(Z,G) \left(Y - \widehat{\mathds{M}}_{\widehat \nu_0}(0,Z,G)\right) + \widehat{\mathds{M}}_{\widehat \nu_0}(0,Z,G) \right\}\right].
\end{align*}
In the above, note that $\widehat{\mathds{M}}_{\widehat \nu_1}(1,Z,G) = \widehat \mu_1(Z,G)$, $\widehat{\mathds{M}}_{\widehat \nu_1}(0,Z,G) = \widehat \mu_{10}(Z,G)$,  and $\widehat{\mathds{M}}_{\widehat \nu_0}(0,Z,G) = \widehat \mu_0(Z,G)$. The nuisance functions $\nu$ and $w$ can both be characterized as solutions to a Fredholm integral equation of the first kind. 
For example, $\nu_1$ can be obtained by solving the empirical integral equation $\widehat{\mathds{M}}_{\nu_1}(1,Z_i,G_i) = \widehat{\mu}_1(Z_i,G_i)$ evaluated at the observed $(Z_i,G_i)$. 
Solving such estimating equations is commonly encountered in nonparametric instrumental variable (IV) models 
\citep{newey2003instrumental,hall2005nonparametric,darolles2011nonparametric} 
and in proximal causal inference 
\citep{miao2018identifying,cui2024semiparametric,tchetgen2024introduction}. When $M$ and $Z$ are discrete, the problem reduces to solving a finite-dimensional system of linear equations. 
When $M$ and $Z$ are continuous, the inverse problem becomes ill-posed and requires regularization. 
Common regularization strategies include imposing a parametric or sieve representation for $w$, applying operator-based regularization methods such as singular value decomposition 
\citep{carrasco2007linear}, spectral cutoff or Tikhonov regularization 
\citep{engl2015regularization}, or adopting semiparametric formulations with implicit regularization 
\citep{cui2023semiparametric}. 
Recently proposed debiasing strategies based on influence functions of modified projection errors 
\citep{ghassami2025debiased} also provide a principled alternative.

Under our identification conditions, the solution for $\nu$ is unique. 
In contrast, the integral equation defining $w$ may admit multiple solutions, 
reflecting the ill-posedness of the operator inversion. 
To stabilize estimation, we recommend incorporating an explicit variance-based penalty 
(e.g., adding the sample variance of the resulting influence function to the objective function) 
when solving for $w$. 
Alternatively, one may directly regularize the norm of $w$. 
This is particularly sensible because $w$ plays a role analogous to inverse probability weights 
in causal inference and missing data settings: large magnitudes of $w$ inflate the variance of the final estimator. 
Norm-based regularization, therefore, improves numerical stability while benefiting efficiency. In the following, we formally establish the asymptotic properties of our proposed one-step estimators under suitable requirements on the nuisance function estimates. 

\begin{theorem}[Multiple robustness of $\widehat \Psi^{\rm NDE}$ and $\widehat \Psi^{\rm NIE}$ ] \label{thm:robust}
Assume that $(\widehat{\mathds{M}}_\nu,\widehat \nu, \widehat \lambda, \widehat w) \xrightarrow[]{p} (\overline{\mathds{M}}_\nu,\bar \nu, \bar \lambda, \bar w)$ with $\widehat{\mathds{M}}_{\widehat \nu} \xrightarrow[]{p} \overline{\mathds{M}}_{\bar \nu}$. Then $\widehat \Psi^{\rm NDE}\xrightarrow[]{p}\rm \psi^{NDE}(P)$ and $\widehat \Psi^{\rm NIE}\xrightarrow[]{p} \rm \psi^{NIE}(P)$ if either:
\begin{enumerate}
    \item $ (\overline{\mathds{M}}_v, \bar \nu) = ({\mathds{M}}_\nu  , \nu)$; or
    \item $(\overline{\mathds{M}}_v, \bar w) = ({\mathds{M}}_\nu, w)$; or
    \item $(\bar \lambda,  \bar w)  =( \lambda,  w)$; or
    \item $(\bar \lambda, \bar{w}^{11}, \bar{w}^{00}, \bar \nu_1)  =( \lambda, w^{11}, w^{00}, \nu_1)$.
\end{enumerate}
\end{theorem}
Theorem~\ref{thm:robust} shows the proposed one-step estimators are multiply robust. In particular, it highlights a symmetry between $\nu$ and $w$. When $\widehat{\mathds{M}}_\nu$ is consistent, it suffices for either $\widehat \nu$ or $\widehat w$ to be consistent, yielding a doubly robust type guarantee. A closer look into scenarios (iii) - (iv) further suggests that, once nuisance parameters $(\lambda, w^{11}, w^{00})$ are consistently estimated, the relevant symmetry is between $\widehat \nu_1$ and $\widehat w^{10}$. This is practically useful since both $\widehat \nu_1$ and $\widehat w^{10}$ are obtained by solving Fredholm integral equations. Consequently, if we posit working models for these two nuisance functions, consistency only requires one of them to be correctly specified. 

\begin{theorem}[Asymptotic behavior of $\widehat \Psi^{\rm NDE}$ and $\widehat \Psi^{\rm NIE}$] \label{thm:asymp}
We let $\lVert f \rVert = \{ \int f(o)^2dP(o)\}^{1/2}$ denote the $L^0_2(P)$-norm. Under Assumptions~\ref{ass: unconfound exposure} to \ref{ass: transportability} and Condition~\ref{cond: med completeness}, if $\widehat{\mathds{M}}_\nu,\widehat \nu, \widehat \lambda,  \widehat w$ all belong to a fixed $P$-Donsker class of functions with probability tending to one, $\lVert D^{\rm NDE}_{\hat{P}} - D_P^{\rm NDE} \rVert  = o_P(1)$ and $\lVert D^{\rm NIE}_{\hat{P}} - D_P^{\rm NIE} \rVert  = o_P(1)$ 
, and 
\begin{enumerate}
    \item $ \max\{\lVert \widehat \nu_0 - \nu_0\rVert, \lVert \widehat \nu_1 - \nu_1\rVert \} \cdot \max\{\lVert \widehat w^{11} -  w^{11}\rVert,\lVert\widehat w^{10} -  w^{10}\rVert,\lVert\widehat w^{00} -  w^{00}\rVert \}  = o_P(n^{-1/2}) $; and
    \item $ \big\{\max\{\lVert \widehat w^{11} -  w^{11}\rVert,\lVert\widehat w^{10} -  w^{10}\rVert,\lVert\widehat w^{00} -  w^{00}\rVert \}  + \lVert \widehat \lambda - \lambda\rVert \big\} \cdot \max \{ \lVert \widehat{\mathds{M}}_{\nu_0} - \mathds{M}_{\nu_0}\rVert,\lVert \widehat{\mathds{M}}_{\nu_1} - \mathds{M}_{\nu_1}\rVert \}=o_P(n^{-1/2})$ .
\end{enumerate}
Then the proposed estimators are asymptotically normal with
\begin{align*}
    &\sqrt{n}(\widehat \Psi^{\rm NDE} - \psi^{\rm NDE}(P)) \rightarrow N(0, \rm var(D^{\rm NDE}_P))\\
    &\sqrt{n}(\widehat \Psi^{\rm NIE} - \psi^{\rm NIE}(P)) \rightarrow N(0, \rm var(D^{\rm NIE}_P)).
\end{align*}
\end{theorem}
Theorem~\ref{thm:asymp} implies that the proposed one-step estimators will be asymptotically linear and normal if the nuisance function estimates converge fast enough but each can be slower than root-n rate. These rates might not be feasible when $Z,G$ is of moderate to high dimension or the extent to which the integration equations defining $w$ and $\nu$ are ill-posed. As a result, in practice, one may posit a working model and maintain doubly robust as stated in Theorem~\ref{thm:robust}. Moreover, $\Psi^{\rm NDE}$ and $\Psi^{\rm NIE}$ attain the semiparametric efficiency bound for estimating the natural direct and indirect effect relative to the model $\mathcal{P}_{\rm comp}$ where $Z$ and $M$ are mutually complete. 

\begin{remark}\label{remark: connect two-sample IV}
    Our framework includes two-sample IV problems as a special case, where $M$ is a binary treatment of interest, $Z$ is a vector of IVs, $G$ is a vector of covariates, and there is no exposure $T$. The target estimand of interest is the average treatment effect $E[Y(M = 1)\mid S = s] - E[Y(M = 0)\mid S = s]$. To place this within our framework, for $\delta \in \{0,1\}$, we introduce a degenerate hypothetical exposure $T^{(\delta)}$ that always equals $1-\delta$. Under the hypothetical intervention that set $T^{(\delta)} = \delta$, we define a potential treatment $M^{(\delta)}(\delta) = \delta$ and set $M^{(\delta)}(1-\delta) = M$. Thus, $M = M^{(\delta)}(T^{(\delta)})$ as $T^{(\delta)} = 1-\delta$ in the observed world. Under this construction, we have  $Y = Y(1 - \delta, M^{(\delta)}(1 - \delta))$ and the potential outcome $Y(M = \delta)$ can be represented as $Y(1 - \delta, M^{(\delta)}(\delta))$. As a result, the potential outcome mean $E[Y( M=\delta)\mid S = s]$ and hence the average treatment effect, falls within the class of targets covered by Theorem \ref{thm:identification}. In this sense, our results extend existing nonparametric IV identification results in \cite{wang2018bounded,chen2025identification,dong2025marginal} to the two-sample setting. Although the hypothetical exposure $T^{(\delta)}$ violates positivity ($P(T^{(\delta)} = \delta) = 0$), this is harmless since the distribution of $M$ given $T^{(\delta)} = \delta, Z, A, X, S$ is known to be a point mass at $\delta$ and need not be estimated from the data and hence the identification for $E[Y(1 - \delta, M^{(\delta)}(\delta))]$ using the identification formula in Theorem \ref{thm:identification} does not require $P(T^{(\delta)} = \delta) \neq 0$. The same idea also extends the valid-IV identification result in Theorem \ref{thm:identification IV set}. The semiparametric theory in the IV setting slightly differs from the current paper because the distribution of $M(\delta)$ is known rather than needing to be estimated. A full semiparametric efficiency analysis for this special case is left to future work.
\end{remark}

\section{Extension: canonical gradients under no unmeasured confounding}
\label{s: extension}
In practice, practitioners may measure a set of candidate IVs for the purpose of bridging. In that case, the dimension of $Z$ may be larger than $M$, making the model $\mathcal{P}$, under which only $M$ is complete in $Z$, the more realistic formulation. While deriving the canonical gradient relative to $\mathcal{P}$ would be ideal, we have not been able to find the form of this gradient; hence, we leave its derivation to future work.  In what follows, we consider the special case when $U = \emptyset$, and denote the corresponding model as $\mathcal{P}^\emptyset$. We first characterize the tangent space of $\mathcal{P}^\emptyset$. 

\begin{lemma}[Characterizing the tangent space of $\mathcal{P}^\emptyset$] \label{lem:tangent_space_noU}
    Under no unmeasured confounding, i.e. $U = \emptyset$, the tangent space of $\mathcal{P}^\emptyset$ at $P \in \mathcal{P}^\emptyset$ takes the following form:
    \begin{align*}
        & \mathcal{T}(P,\mathcal{P}^\emptyset) = \Big\{o \mapsto (1-s)\left\{E_P[h_Y(Y,M,T,G) \mid y,t,z,g] +E_P[h_M(M,T,Z,G) \mid y,t,z,g]\right\}  \\
        & \hspace{4em} + s\cdot h_M(m,t,z,g)  + h_T(t,z,g,s) + h_Z(z,g,s) + h_G(g,s) + h_S(s)\\
        & \hspace{4em}: h_Y\in L^0_2(P_{Y\mid M,T,G}),h_M\in L^0_2(P_{M\mid T,Z,G,S=1}), \\
        & \hspace{5em} h_T\in L^0_2(P_{T\mid Z,G,S}), h_Z\in L^0_2(P_{Z\mid G,S}), h_G\in L^0_2(P_{G\mid S}), h_S\in L^0_2(P_{S})\Big\},
    \end{align*}
    where in the above, we take the expectations under $P$ as the joint conditional distribution $P_{Y,M\mid T,Z,G}$ can be identified from the observed distribution when $U = \emptyset$ \citep{li2024calibrated}.
\end{lemma}
In the absence of unmeasured confounding and when only $M$ is complete in $Z$, we now see how the conditional independence involving the IVs and latent alignment in $(Y,M)$ across data sources restrict the tangent space. In particular, the subspace corresponding to $P_{Y\mid T,Z,G,S=0}$ is no longer the entire Hilbert space $L^0_2(P_{Y\mid T,Z,G,S=0})$. Instead, it now consists of two components, one of which is shared with the $S=1$ source along the direction of the mediator $h_M$. 

\sloppy For ease of notation, we denote $c^{11}(y,t,z,g)  = t w^{11}(z,g)\{y-  \mu_1(z,g)\}/P(S=0)$, $c^{00}(y,t,z,g)  = (1-t) w^{00}(z,g)\{y-  \mu_0(z,g)\}/P(S=0)$, and $c^{10}(y,t,z,g)  = t w^{10}(z,g)\{y-  \mu_1(z,g)\}/P(S=0)  - \lambda(t,z,g)\{ E_P[\nu_1(M,G) \mid y,t,z,g] - \mathds{M}_{\nu_1}(t,z,g)\}  \{(1-t) w^{00}(z,g) - t w^{10}(z,g)\}/P(S=1)$. The canonical gradients of $\psi^{11}$, $\psi^{10}$, and $\psi^{00}$ admit the common representation below.

\begin{theorem}[The canonical gradients of $\psi^{\rm NDE}$ and $\psi^{\rm NIE}$ relative to $\mathcal{P}^\emptyset$] \label{thm: EIF_noU}
    Under Assumptions~\ref{ass: unconfound exposure} to \ref{ass: transportability} and Condition~\ref{cond: med completeness}, for each $k \in \{11,10,00\}$, the canonical gradient of $\psi^{k}$ relative to $\mathcal{P}^\emptyset$ at $P$ is
    $$\tilde{D}^{k}_{P}(o) = D^{k}_{P}(o)  - (\mathcal{A}_P\tilde{e}^k)(o),$$
    \sloppy where $(\mathcal{A}_Pe)(o) :=  \mathds{1}(s=0) \left(e(y,t,z,g) - E_P[e(Y,T,Z,G) \mid t,z,g]\right)- \mathds{1}(s=1) P(S=0)/P(S=1)\lambda(t,z,g) \left(E_P\left[ e(Y,T,Z,G) \mid m,t,z,g\right] - E_P[e(Y,T,Z,G) \mid t,z,g]\right)$, and $\tilde{e}^k(y,t,z,g)$ is the solution of the optimization problem
    \begin{align*}
        \mathop{\arg\min}_{
        \substack{
        \text{\tiny $e^k$: function of}\\ \text{\tiny $(y, t, z, g)$}}} \Bigg\{\inf_{\substack{
        \text{\tiny $\xi^k$: function of }\\ \text{\tiny $(y, m, t, g)$}}}
        \Bigg\{ E_P[\mathcal{L}_P(O;c^k,\xi^k,e^k)]
        \Bigg\}
        \Bigg\},
    \end{align*}
    \sloppy where $\mathcal{L}_P(o;c,\xi,e) =\{c(y,t,z,g) - E_P[\xi(Y,M,T,G)\mid y,t,z,g] - e(y,t,z,g) + E_P[e(Y,T,Z,G)\mid t,z,g] - \frac{P(S = 0)}{P(S = 1)}\lambda(t,z,g)E_P[E_P[(e(Y,T,Z,G) - E_P[e(Y,T,Z,G)\mid T,Z,G])\mid M,T,Z,G]\mid y,t,z,g] \}^{2} + E_P[e(Y,T,Z,G)\mid y,m,t,g]^{2} + E_P[\xi(Y,M,T,G)\mid m,t,g]^{2}$.
    The canonical gradients for $\psi^{\rm NDE}$ and $\psi^{\rm NIE}$ relative to $\mathcal{P}^{\emptyset}$ at $P$ are $\tilde{D}^{\rm NDE}_P(o) = \tilde{D}^{10}_{P}(o) - \tilde{D}^{00}_{P}(o)$ and $\tilde{D}^{\rm NIE}_P(o)= \tilde{D}^{11}_{P}(o) - \tilde{D}^{10}_{P}(o)$.
\end{theorem}
Compared with Theorem~\ref{thm: mutual complete} where mediator-only sample ($S=1$) does not improve efficiency for estimating $\psi^{11}(P)$ and $\psi^{00}(P)$, Theorem~\ref{thm: EIF_noU} shows that such efficiency gains can arise when $Z$ is not complete in $M$. This highlights the role of mutual completeness in our setting. Under the considered causal mechanism, $Z$ affects $Y$ only through the mediator $M$. When $Z$ is not complete in $M$, inverse-propensity score weighting -- which balances the distribution of $(Z,G)$ across groups -- can be overly restrictive. Additional efficiency gain can be achieved by adopting alternative weights that target only on the component of the $Z$ distribution that changes the induced mixture over $M$.

\section{Simulation}
\label{s:simulation}

\subsection{Simulations for estimating NDE and NIE}
\label{sec:simulation}

We examined the finite-sample performance of the proposed estimators under two instrument structures: (1) a single-instrument $Z \in R$ where $Z$ and $M$ are mutually complete, and (2) a two-instrument $Z \in R^2$ where only $M$ is complete in $Z$. For both settings, we let $S \sim \mathrm{Bernoulli}(0.4)$ and $T \mid S=s \sim \mathrm{Bernoulli}\{0.4(1-s) + 0.5s\}$. In the first setting, we generated $Z \mid T=t,S=s \sim N\{\mu(t,s),1\},$ where $\mu(0,0)=0.6$, $\mu(1,0)=0.5$, $\mu(0,1)=0.3$, and $\mu(1,1)=0.4$. Let $U$, $\varepsilon_M$, and $\varepsilon_Y$ be three standard normal variables where \(\varepsilon_M\) and \(\varepsilon_Y\) are independent of each other and of \((T,S,Z,U)\). Conditional on $T, Z, U$, the mediator was generated from $ M=\alpha_{0t}+\alpha_{Zt} Z+U+\varepsilon_M$, with $(\alpha_{00},\alpha_{Z0})=(-0.4,0.4)$ and $(\alpha_{01},\alpha_{Z1})=(-0.1,1.5)$. The outcome was generated from $Y=\beta_{0t}+\beta_{Mt} M+U+\varepsilon_Y$, $\varepsilon_Y \sim N(0,1)$, where $(\beta_{00},\beta_{M0})=(0.5,-0.2)$ and $(\beta_{01},\beta_{M1})=(0.5,-0.8)$. In the second setting, the instrument \(Z=(Z_1,Z_2)\) is two-dimensional with
$Z_j \mid T=t,S=s \sim N\{\mu(t,s),1\}$ for $j = 1, 2$,
where \(\mu(t,s)\) is the same conditional mean function as that in the setting with a single IV. Conditional on $T, Z, U$ the mediator model was
$
M=\alpha_{0t}+\alpha_{Zt,1}Z_1+\alpha_{Zt,2}Z_2+U+\varepsilon_M$,
with
$(\alpha_{00},\alpha_{Z0,1},\alpha_{Z0,2})=(-0.1,0.4,0.6)$ and
$(\alpha_{01},\alpha_{Z1,1},\alpha_{Z1,2})=(0.4,0.7,0.8)$.
The outcome followed the same generating process as in the first setting. Throughout, the target estimand was the natural indirect effect in source $S = 0$. 

For each setting, we report the proposed one-step estimators with nuisance components estimated by correctly specified parametric models  (\texttt{Correct}) together with four estimators with different combinations of model misspecification listed in Theorem \ref{thm:robust} that still admit consistent estimation (\texttt{Mis 1}--\texttt{Mis 4}). For the two-instrument setting, we examined one-step estimators that uses an aggregated instrument $\widetilde Z_a=aZ_1+(1-a)Z_2$ for some $a\in[0,1]$, where \(a\) is selected adaptively by minimizing an estimated variance criterion. We also included  a \texttt{Single IV} estimator that uses only $Z_2$ with correctly specified nuisance models. In each Monte Carlo replicate, we generated \(n=1000\) observations and repeated the experiment over 1000 replicates. Additional information on how the models are misspecified can be found in Supplementary Section S8.

Table~\ref{tab: sim cont} reports the bias, SE, and coverage of the proposed estimators. All estimators exhibit negligible bias.  In the complete setting, the \texttt{Correct} estimator is the most efficient, whereas the misspecified estimators exhibit a noticeable loss of efficiency, with the magnitude of the loss depending on the misspecification pattern. In the non-complete setting, the advantage of combining the two instrument components is substantial: the proposed \texttt{Correct} estimator has a much smaller SE than the \texttt{Single IV} estimator, demonstrating clear efficiency gain from adaptive instrument aggregation. For CI coverage, both CI-IF and CI-Boot generally attain coverage rates close to or slightly above the nominal level. The bootstrap-based intervals are overall more stable across different misspecification patterns, while the influence-function-based intervals tend to be somewhat conservative in several misspecified cases. Overall, the simulation results suggest that the proposed estimator remains stable under the misspecification schemes considered, and that the bootstrap-based variance estimation provides more reliable inference than the plug-in influence-function-based variance estimator.

\begin{table}[h]
\caption{Bias, SE and the coverage rate of the CIs for different estimators with a continuous mediator.}\label{tab: sim cont}
    \centering
    \begin{tabular}{llcccc}
    \toprule
    Setting &Estimator & Bias & SE & CI-IF & CI-Boot \\
    \midrule
    \multirow{5}{*}{Complete} &\texttt{Correct} & 0.00 & 0.45& 93.6\%& 97.7\%\\
    &\texttt{Mis 1} & 0.0& 0.60& 97.2\%&  98.4\%\\
    &\texttt{Mis 2} & -0.03& 0.77&  100.0\%&  95.7\%\\
    &\texttt{Mis 3} & 0.00& 0.63& 98.8\%&  99.3\%\\
    &\texttt{Mis 4} & -0.03& 1.03& 99.7\%&  97.3\%\\
    \midrule
    \multirow{6}{*}{Noncomplete} &\texttt{Correct} & 0.00& 0.16& 94.7\%& 95.3\%\\
    &\texttt{Mis 1} &  -0.01& 0.20&  98.2\%& 96.6\%\\
    &\texttt{Mis 2} &  0.00& 0.17&  98.9\%& 95.0\%\\
    &\texttt{Mis 3} &  0.00& 0.23& 99.4\%& 97.5\%\\
    &\texttt{Mis 4} & 0.00& 0.26& 98.5\%& 96.6\%\\
    &\texttt{Single IV} & 0.00& 0.59& 94.2\%& 94.1\%\\
    \bottomrule
    \end{tabular}
\end{table}

\subsection{Simulations for the IV selection procedure}
We also evaluated the numerical performance of the IV selection procedure proposed in Supplementary Material Section S7. The simulation setting was the same except that an additional invalid IV $Z_3 \sim N(0, 1)$ is included and $M$ and $Y$ were generated from $
M=\alpha_{0t}+\alpha_{Zt,1}Z_1+\alpha_{Zt,2}Z_2+Z_3+U+\varepsilon_M$ and $
Y=\beta_{0t}+\beta_{Mt} M + \nu Z_3+U+\varepsilon_Y$,
where the parameter \(\nu\) controls the degree of violation of the IV assumption. We considered two values, \(\nu=1\) and \(\nu=2\), representing moderate and strong invalidity, respectively. Table~\ref{tab:IV sel} reports the probability that the procedure in Supplementary Material Section S7 correctly identifies \(Z_3\) as the invalid IV under different combinations of \(n\) and \(\nu\).

\begin{table}[h]
    \centering
    \caption{Probability that the IV selection algorithm in Supplementary Material Section S7 correctly identifies the invalid IV}
    \label{tab:IV sel}
    \begin{tabular}{cccc}
    \toprule
         $n$& $ 500$ &$1000$ & $2000$  \\
    \midrule
         $\nu=1$&97.0\% & 99.7\% & 100\%  \\
         $\nu=2$& 100\%& 100\% &100\%\\
    \bottomrule
    \end{tabular}
\end{table}

The results show that the proposed procedure performs well across all settings considered. When \(\nu=1\), the probability of correctly identifying \(Z_3\) increases from \(97.0\%\) at \(n=500\) to \(100\%\) at \(n=2000\), indicating that the procedure becomes increasingly reliable as the sample size grows. When the violation is stronger, with \(\nu=2\), the procedure identifies \(Z_3\) correctly in essentially every replicate for all three sample sizes.

\section{Data Application}
\label{s:data}
Dementia and Alzheimer's disease (AD) are complex neurodegenerative disorders influenced by both environmental and genetic factors. Genome-wide association studies (GWAS) have identified numerous genetic variants associated with dementia risk \citep{lambert2013meta,kunkle2019genetic}. Many of these variants lie in non-coding regions of the genome, suggesting that they may influence disease susceptibility through regulatory mechanisms, such as changes in gene expression \citep{hollingworth2011common,reus2020gene}. Consistent with this view, transcriptomic studies have documented widespread gene-expression changes in AD brain tissue, and peripheral-blood expression profiles have also been associated with clinical severity and progression from mild cognitive impairment to dementia \citep{roed2013prediction, wan2020meta}. Together, these findings suggest that gene expression may serve as an important intermediate mechanism linking genetic variation to dementia progression \citep{zhao2026systems}. Quantifying the extent to which genetic effects are mediated through different gene-expression pathways may therefore provide insight into the biological mechanisms underlying dementia progression.

We investigate the extent to which the effect of the single nucleotide polymorphism (SNP) rs610932 on dementia risk is mediated through the expression of two selected genes, namely \textit{PTK2B}  and \textit{CD33}. rs610932 lies in the MS4A locus, near MS4A6A, and has been repeatedly associated with Alzheimer’s disease risk in genome-wide association studies and subsequent meta-analyses \citep{hollingworth2011common}. We selected CD33 and PTK2B expression as candidate mediators as the MS4A locus has been linked to microglial-state regulation and soluble TREM2 \citep{deming2019ms4a}, while CD33 \citep{griciuc2013alzheimer} and PTK2B \citep{guo2023microglia} are both implicated in AD-relevant microglial and innate-immune pathways, making them plausible downstream mediators.

We analyze the observational database from the  Alzheimer’s Disease Neuroimaging Initiative (ADNI). ADNI is a longitudinal, multicenter observational study launched in 2004 and continued through successive phases, including ADNI-1, ADNI-GO, ADNI-2, ADNI-3, and ADNI-4. Its goal is to develop and validate biomarkers for Alzheimer’s disease, with repeated imaging, clinical, cognitive, and genetic assessments collected over more than 20 years of follow-up. Importantly, gene-expression data were not available in ADNI-1. Transcriptomic data became available only in later phases, primarily for selected ADNI-GO/2 participants and onwards. As a result, dementia outcomes, SNPs, and gene-expression measurements were never jointly observed for the ADNI-1 cohort. 

To investigate the effects of rs610932 on dementia risk through different gene-expression pathways for the ADNI-1 cohort, we combine data sources from both ADNI-1 and ADNI-GO/2 phases and a descriptive table can be found at Supplementary Section S9. We restrict to all individuals with an observed diagnosis of dementia, while censored individuals were included only if their follow-up time exceeded 7 years. We define the outcome $Y$ to be dementia diagnosis at year 7 since enrollment. The analysis cohort consists of $n_0=328$ ADNI-1 participants with $(X,D,Y)$ available, and $n_1=352$ ADNI-GO/2 participants with $(X,D,M)$ available. The exposure of interest $T$ is SNP rs610932, coded as indicator of carrying one or more minor allele.  The mediator vector $M$ consists of the expression levels of \textit{PTK2B} and \textit{CD33}. Baseline covariates $X$ include age, sex, years of education, APOE genotype, diagnosis of mild cognitive impairment at enrollment, and mini-mental state examination score. For each mediator of interest, we first selected the top 50 most correlated SNPs based on eQTL summaries \citep{gtex2020atlas}. We then identified SNPs overlapping with those available in the ADNI study, yielding $\ccC={rs3735759,rs1879189,rs9644121}$ for PTK2B and $\ccC={rs3826656,rs273638,rs273634}$ for CD33. We next applied the proposed IV selection algorithm to construct the instrument set $\cZ$ for each mediator. Notably, the algorithm recovered the same SNP sets.

\begin{figure}[H]
    \centering
    \includegraphics[width=0.7\linewidth]{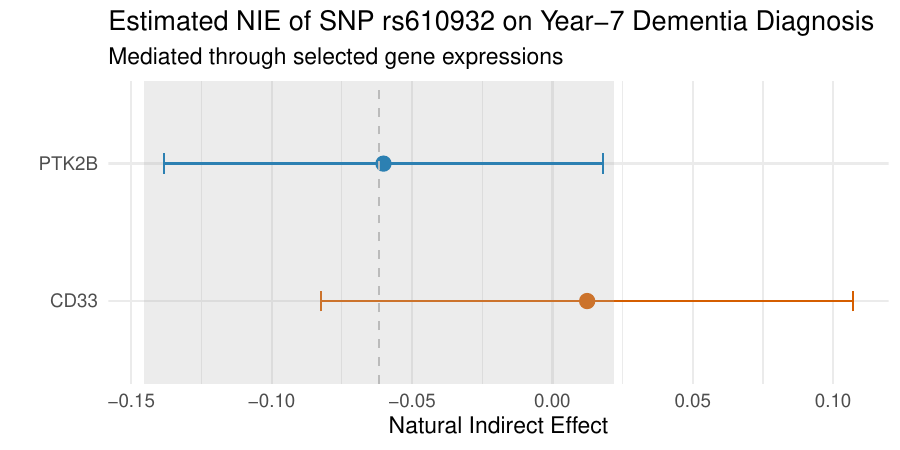}
    \caption{Estimated natural indirect effects mediated by selected gene expressions. The estimated average total effect is $-0.06$ with $95\%$ CI $[-0.15, 0
    .02]$ indicated by the gray region.}
    \label{fig:NIE}
\end{figure}

The estimated total average effect of rs610932 on 7-year dementia risk is estimated to be $-0.06$ (confidence interval $[-0.15, 0.02]$). The estimated natural indirect effects are plotted in
Figure~\ref{fig:NIE}. Using the corresponding selected IVs as instruments, we found a heterogeneous but biologically coherent pattern of indirect effects across the candidate mediators: the estimated NIEs were negative and substantial relative to the size of the total effect for PTK2B, and close to null for CD33, although all confidence intervals remained wide and crossed zero, so these results should be viewed as suggestive rather than definitive. These findings suggest that the protective effect of the rs610932 minor allele may operate through altered \textit{PTK2B} expression and, more broadly, through downstream perturbation of a coordinated microglial immune-expression program. In addition, we have conducted a sensitivity analysis with respect to the choice of $\cZ$ and defer reader's attention to Supplementary Section S9.

\section*{Acknowledgments}
We gratefully acknowledge access to data from the Alzheimer’s Disease Neuroimaging Initiative (ADNI). These data were used under approved data-use agreements. We thank Daxuan Deng for generously sharing their code for processing the ADNI datasets.

\bibliography{MedFusion.bib}

\end{document}